\documentstyle{pss}
\input epsf.tex
\begin{document}

\title{ Orbital Ordering and Orbital Fluctuations \\ 
        in Transition Metal Oxides }

\author{{\sc Andrzej M. Ole\'s}\footnote{) E-mail: A.M.Oles@fkf.mpg.de } ) }
\address{Marian Smoluchowski Institute of Physics, Jagellonian 
 University,\\ 
 Reymonta 4, PL-30059 Krak\'ow, Poland \\
 Max-Planck-Institut f\"ur Festk\"orperforschung, Heisenbergstrasse 1, \\
 D-70569 Stuttgart, Federal Republic of Germany}
\submitted{1 July, 2002} 
\maketitle
\hspace{9mm} 

\begin{abstract}
We summarize some characteristic features of the frustrated magnetic 
interactions in spin-orbital models adequate for cubic transition metal 
oxides with orbital degeneracy. A generic tendency towards dimerization, 
found already in the degenerate Hubbard model, is confirmed for $t_{2g}$ 
but not for $e_g$ systems. In the $t_{2g}$ case the quantum orbital 
fluctuations are more pronounced and contribute to a stronger 
competition between different magnetic and orbital states. Therefore the 
orbital liquid states exist in some undoped $t_{2g}$ systems, while in 
the manganites such states can be triggered only by doping.\\
Journal reference: A. M. Ole\'s, Phys. Stat. Sol. (b) {\bf 236}, 281 (2003).
\end{abstract}

The physical properties of transition metal oxides are dominated by 
large on-site Coulomb interactions $\propto U$ which suppress charge 
fluctuations. Therefore, such systems are either Mott or
charge-transfer insulators, and the metallic behavior might occur only 
as a consequence of doping. Here we will discuss first the undoped 
systems with localized $d$ electrons which interact by effective 
superexchange (SE) interactions. An interesting situation occurs when 
$d$ electrons occupy partly degenerate orbital states, and one has to 
consider {\it orbital degrees of freedom\/} in the SE at equal footing 
with electron spins \cite{Tok00}. Competition between different states 
is then possible, holes may couple to orbital excitations \cite{Zaa93}, 
and the quantum effects are enhanced already in undoped systems 
\cite{Fei97}. The first models of SE in such situations were proposed 
almost three decades ago \cite{Kug73}, either by considering the 
degenerate Hubbard model \cite{Cla75,Ina75}, or for realistic situations 
encountered in cuprates (KCuF$_3$ and K$_2$CuF$_4$) and in V$_2$O$_3$ 
\cite{Cas78}. Then it was realized that the SE which is usually 
antiferromagnetic (AF) might become ferromagnetic (FM) when Hund's 
exchange interaction $J_H$ is finite, but only in recent years the 
phenomena which originate from {\it the orbital physics\/} are 
investigated in a more systematic way.

The SE which involves the orbital degrees of freedom is described by the 
so-called {\it spin-orbital models\/} \cite{Ole01}, and is typically 
highly frustrated on a cubic lattice where it might even lead to 
the collapse of magnetic long-range order by strong spin or orbital 
fluctuations \cite{Fei97}. However, in real $e_g$ systems such quantum 
phenomena are usually quenched by finite $J_H$ which induces a 
structural phase transition and thus helps to stabilize a particular 
ordering of occupied orbitals which supports $A$-type AF order, 
as observed when degenerate orbitals are filled either by one hole 
(KCuF$_3$) \cite{Ole00}, or by one electron (LaMnO$_3$) \cite{Fei99}. 
The coupling to the lattice due to the Jahn-Teller (JT) effect also 
helps to stabilize the orbital ordering, and quantitative models of the 
structural transition have to include both these effects \cite{Fei99}.  

\begin{figure}[t!]
\begin{tindent}
\epsfxsize=10cm
\epsffile{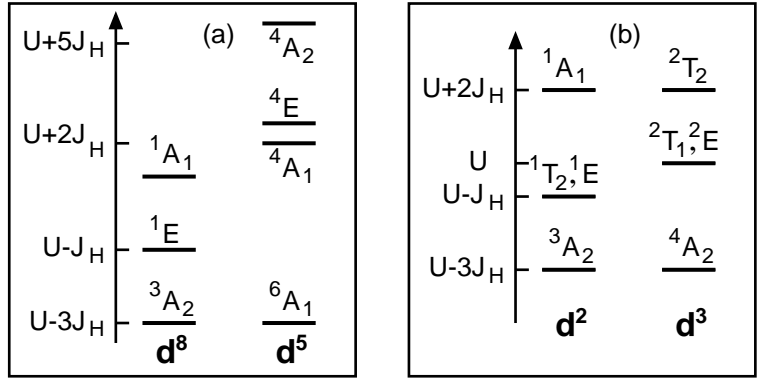}
\end{tindent}
\caption{
 Excitation spectra in cubic transition metal oxides for: 
 (a) $e_g$ systems: Cu$^{3+}$ ($d^8$) and Mn$^{2+}$ ($d^5$) ions;
 (b) $t_{2g}$ systems: Ti$^{2+}$ ($d^2$) and  V$^{2+}$ ($d^3$) ions
 \protect\cite{Ole01}.}
\label{fig:ex}
\end{figure}

The essential feature of the SE described by spin-orbital models is the 
frustration of magnetic interactions: the FM terms occur next to the AF 
ones, and it depends on the physical parameters which interactions 
finally win and stabilize a given type of magnetic order. The simplest 
spin-orbital model which illustrates this physics may be derived for the 
density of one electron per site ($n=1$) in a doubly degenerate Hubbard 
model. We assume that the hopping is isotropic and diagonal between two 
orbitals $\alpha$ and $\beta$ at sites $i$ and $j$, and include only the 
on-site interaction elements: Coulomb $U$ and Hund's exchange $J_H$, 
\begin{eqnarray}
H\!\!&=&\!\!-t\sum_{ij\mu\sigma}
   a^{\dagger}_{i\mu\sigma}a^{}_{j\mu\sigma} 
  +U\sum_{i\mu}n^{}_{i\mu\uparrow}n^{}_{i\mu\downarrow} 
  +(U-\frac{5}{2}J_H)\sum_{i}n^{}_{i\alpha}n^{}_{i\beta}     \nonumber \\
 &-&\!\!2J_H\sum_{i}\vec{S}^{}_{i\alpha}\vec{S}^{}_{i\beta}+J_H\sum_{i}
   (a^{\dagger}_{i\alpha  \uparrow}a^{\dagger}_{i\alpha\downarrow}
    a^{       }_{i\beta \downarrow}a^{       }_{i\beta   \uparrow}
   +a^{\dagger}_{i\beta   \uparrow}a^{\dagger}_{i\beta \downarrow}
    a^{       }_{i\alpha\downarrow}a^{       }_{i\alpha  \uparrow}),
\label{hub}
\end{eqnarray}
with the spin,
$\{S_{i\mu}^+,S_{i\mu}^-,S_{i\mu}^z\}=
 \{a^{\dagger}_{i\mu  \uparrow}a^{}_{i\mu\downarrow},
   a^{\dagger}_{i\mu\downarrow}a^{}_{i\mu  \uparrow},
  (n_{i\mu\uparrow}-n_{i\mu\downarrow})/2\}$, and density operators, 
$n_{i\mu}=n_{i\mu\uparrow}+n_{i\mu\downarrow}$, at orbital 
$\mu=\alpha,\beta$ defined in the usual way. The interactions are 
rotationally invariant in the orbital space \cite{Ole83}. If $U\gg t$, 
the electrons localize and the low-energy physics is described by the 
SE interactions which follow from the virtual 
$d_i^1d_j^1\rightleftharpoons d_i^0d_j^2$ processes on the bonds 
$\langle ij\rangle$. They lead either a high-spin $^3A_2$ state, or to
a double occupancy in either orbital which has to be subsequently 
projected onto two low-spin $^1E$ and $^1A_1$ states. The excitation 
spectum is equidistant, as for $d^8$ ions in the cuprates [Fig. 
\ref{fig:ex}(a)], with the excitation energies derived from Eq. 
(\ref{hub}):
$\varepsilon(^3\!A_2)=U-3J_H$, 
$\varepsilon(^1\!E)  =U- J_H$, and  
$\varepsilon(^1\!A_1)=U+ J_H$  \cite{Ole00}. The SE Hamiltonian derived 
from Eq. (\ref{soiso}) takes the form,
\begin{eqnarray}
\label{soiso}
H_{I}\!\!&=&\!\!Jr_1\sum_{\langle ij\rangle}
     \left(\vec{S}_i\cdot\vec{S}_j+\frac{3}{4}\right)
     \left(\vec{T}_i\cdot\vec{T}_j-\frac{1}{4}\right)
     +\frac{1}{2}J\sum_{\langle ij\rangle}
     \left(\vec{S}_i\cdot\vec{S}_j-\frac{1}{4}\right)      \nonumber \\
&\times &\!\!\left[r_2(1\!+\!T_i^+T_j^-\!+\!T_i^-T_j^+)
               +r_3(\frac{1}{2}+2T_i^zT_j^z)
           +(r_2-r_3)(T_i^+T_j^+\!+\!T_i^-T_j^-)\right],     
\end{eqnarray}
where $J=4t^4/U$ is the energy unit for the SE interaction, and the 
coefficients $r_1=1/(1-3\eta)$, $r_2=1/(1-\eta)$, $r_3=1/(1+\eta)$ 
follow from the above charge excitations, where $\eta=J_H/U$. Similar 
to spin, the pseudospin operators are:  
$\{T_{i}^+,T_{i}^-,T_{i}^z\}=
 \{\sum_{\sigma}a^{\dagger}_{i\alpha\sigma}a^{}_{i\beta \sigma},
   \sum_{\sigma}a^{\dagger}_{i\beta \sigma}a^{}_{i\alpha\sigma},
   (n_{i\alpha}-n_{i\beta})/2\}$. It is
important to use the accurate form of the electron-electron interactions
\cite{Cla75,Ole00}, and for this reason some early work led to 
inaccurate expressions \cite{Kug73,Ina75}. Note that spin interactions 
have SU(2) symmetry, while the orbital interactions are anisotropic. The 
first term is simple and follows from the excitations of spin triplet 
and interorbital singlet state. The low-spin terms $\propto r_{2(3)}$ 
are more involved and include not only orbital-flip processes, but also 
pair hopping terms $\propto (T_i^+T_j^+\!+\!T_i^-T_j^-)$. This 
demonstrates that the anisotropy in the orbital sector is a feature 
which follows from the multiplet spectra of transition metal ions 
\cite{Gri71}, where the orbital triplet state never occurs at $J_H>0$.    

The model (\ref{soiso}) simplifies in the limit of $J_H\to 0$, and 
represents a superposition of excitations which involve either spin 
triplet and orbital singlet, or spin singlet and orbital triplet, 
\begin{equation}
\label{su4}
H_{I}\!=J\sum_{\langle ij\rangle}\left[
     \left(\vec{S}_i\cdot\vec{S}_j+\frac{3}{4}\right)
     \left(\vec{T}_i\cdot\vec{T}_j-\frac{1}{4}\right)
    +\left(\vec{S}_i\cdot\vec{S}_j-\frac{1}{4}\right)     
     \left(\vec{T}_i\cdot\vec{T}_j+\frac{3}{4}\right)\right],
\end{equation}
which is just a different way of writing the SU(4) symmetric 
spin-orbital model \cite{Li98}. In this case the spin and orbital 
correlations obey full SU(4) symmetry, and the correlations functions:
$\langle \vec{S}_i\cdot\vec{S}_j\rangle$, 
$\langle \vec{T}_i\cdot\vec{T}_j\rangle$, 
$\frac{4}{3}\langle (\vec{S}_i\cdot\vec{S}_j)
                   (\vec{T}_i\cdot\vec{T}_j)\rangle$, 
are all identical \cite{Fri99}. This condition is violated when
the mean-field approximation (MFA) is used and the spin and orbital
variables are decoupled, so the results of the MFA might be unreliable.  

\begin{figure}[t!]
\begin{tindent}
\epsfxsize=12cm
\epsffile{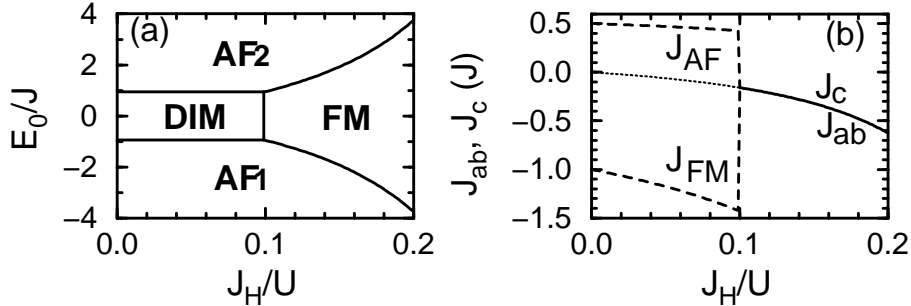}
\end{tindent}
\caption{
(a) Mean-field phase diagram in $(J_H,E_0)$ plane of the isotropic 
spin-orbital model (\protect\ref{soiso}) with four different magnetic 
phases: DIM, FM, AF1 and AF2 (with either $\alpha$ or $\beta$ orbitals 
occupied); (b) exchange constants $J_{ab}$ and $J_c$ 
for increasing $J_H$ at $E_0=0$ as obtained in the DIM phase ($J_{FM}$ 
and $J_{AF}$) and in the FM phase ($J_{ab}=J_c$). }
\label{fig:iso}
\end{figure}

Although some qualitative arguments were given, the classical phase 
diagram of the spin-orbital model (\ref{soiso}) was not investigated 
before. We include the orbital splitting at every site,
$\sim E_0(n_{i\alpha}-n_{i\beta})/2$, and compare the energies of four 
different three-dimensional (3D) phases: 
  $(i)$ AF long-range order (LRO) with either $\alpha$ or $\beta$ 
        orbital occupied at every site, 
 $(ii)$ FM phase with alternating $\alpha/\beta$ orbitals on two 
        sublattices, and
$(iii)$ a dimer phase (DIM) characterized by orbital valence bond (OVB) 
        states, with orbital singlets at every second bond along $c$ 
	axis (or any other, as the present problem is isotropic). 
When the orbital singlet is formed on a single bond, the energy gain 
due to the first term in Eq. (\ref{soiso}) is maximized and the FM 
interaction follows. This leads at small $J_H$ to a DIM state in the 
model for vanadates \cite{She02}, as we will discuss below. On the 
contrary, the orbitals are uncorrelated at all other bonds, the AF 
terms win as long as the Hund's interaction is weak (Fig. 
\ref{fig:iso}). The AF states: AF$1$ and AF$2$ are stabilized 
by the orbital splitting $E_0$ which has to counterbalance the energy 
gains on the FM bonds. Of course, it is hard to imagine that the DIM 
state with ordered orbital singlets might be realized as such, but the 
alternation of FM/AF bonds is plausible, so the present phase diagram 
should rather be viewed as demonstrating a generic competition between 
different signs of the SE interactions. It shows that one may indeed 
expect enhanced quantum fluctuations close to the orbital degeneracy 
when $J_H$ is small \cite{Fei97}.     

The simplest {\it realistic\/} spin-orbital model describes $d^9$ ions 
interacting on a cubic lattice, as in KCuF$_3$. The interactions are the 
same as in Eq. (\ref{hub}), but the hopping term is now nondiagonal and 
allows for orbital excitations \cite{Zaa93}. In the limit of $U\gg t$ 
the charge excitations $d_i^9d_j^9\rightleftharpoons d_i^8d_j^{10}$ lead 
again to the same excited states as above, with their energies [Fig. 
\ref{fig:ex}(a)] reproducing the exact spectrum of $d^8$ ions 
\cite{Gri71}. We define the SE $J=4t^2/U$ by the largest hopping element 
$t$ between two $|z\rangle=|3z^2-r^2\rangle$ orbitals along the $c$ 
axis, and one finds,
\begin{equation}
\label{socu}
{\cal H}(d^9)=\frac{1}{4}J\sum_{\gamma}
                          \sum_{\langle ij\rangle\parallel\gamma}
    \Big[\Big({\vec S}_i\cdot {\vec S}_j+\frac{1}{4}\Big)
    {\hat J}_{ij}^{(\gamma)}(d^9) + {\hat K}_{ij}^{(\gamma)}(d^9)\Big],
\end{equation}
where $\gamma=a,b,c$, and ${\vec S}_i$ are spin $S=1/2$ operators. The 
operator expressions:
\begin{eqnarray}
\label{j9}
{\hat J}_{ij}^{(\gamma)}(d^9)&\!=&\!
(2+\eta r_2-\eta r_3){\cal P}_{\langle ij\rangle}^{\zeta\zeta}
      -\eta(3r_1-r_2){\cal P}_{\langle ij\rangle}^{\zeta\xi},        \\
\label{k9}
{\hat K}_{ij}^{(\gamma)}(d^9)&\!=&\!
-[1+\eta(3r_1+r_2)/2]{\cal P}_{\langle ij\rangle}^{\zeta\xi}
-[1+\eta( r_2-r_3)/2]{\cal P}_{\langle ij\rangle}^{\zeta\zeta},
\end{eqnarray}
describe spin and orbital SE, and the coefficients $r_i$ are defined as
in Eq. (2). The operators:
\begin{eqnarray}
\label{porbit}
{\cal P}_{\langle ij\rangle}^{\zeta\xi}&\!=&\!
 (1/2+\tau^{\gamma}_i)(1/2-\tau^{\gamma}_j)
+(1/2-\tau^{\gamma}_i)(1/2+\tau^{\gamma}_j),       \\
{\cal P}_{\langle ij\rangle}^{\zeta\zeta}&\!=&\!
2(1/2-\tau^{\gamma}_i)(1/2-\tau^{\gamma}_j),
\end{eqnarray}
project on the orbital states, being either parallel to the bond 
$\langle ij\rangle$ direction on one site
($P_{i\zeta}=1/2-\tau^{\gamma}_i$) and perpendicular on the other
($P_{j  \xi}=1/2+\tau^{\gamma}_j$) one, or parallel on both sites. 
They are represented by the orbital operators $\tau^{\gamma}_i$ for 
the three cubic axes:
\begin{equation}
\label{orbop}
\tau^{a(b)}_i = ( -\sigma^z_i\pm\sqrt{3}\sigma^x_i )/4, \hskip .7cm
\tau^c_i = \sigma^z_i/2,
\end{equation}
where the $\sigma$'s are Pauli matrices acting on:
$|x\rangle ={\scriptsize\left( \begin{array}{c} 1\\ 0\end{array}\right)},\;
 |z\rangle ={\scriptsize\left( \begin{array}{c} 0\\ 1\end{array}\right)}$,
which transform as $|x\rangle \propto x^2-y^2$ and
$|z\rangle \propto (3z^2-r^2)/\sqrt{3}$.

In LaMnO$_3$ the SE is more involved and couples {\it total spins\/} 
$S=2$ at the Mn$^{3+}$ ions. It originates from the charge excitations,
$d_i^4d_j^4\rightleftharpoons d_i^3d_j^5$ \cite{Fei99}. The $e_g$ part,
following from $d_i^4d_j^4\rightleftharpoons d_i^3(t_{2g}^3)
d_j^5(t_{2g}^3e_g^2)$ processes, involves again FM terms due to the 
high-spin $^6A_1$ state, and three AF terms due to the low-spin states: 
$^4A_1$, $^4E$, and $^4A_2$ [Fig. \ref{fig:ex}(a)], and has analogous
orbital dependence as in the cuprate case. In contrast,
the $t_{2g}$ part follows only from low-spin excitations 
$d_i^4d_j^4\rightleftharpoons d_i^3(t_{2g}^3)d_j^5(t_{2g}^4e_g)$
and is therefore AF and almost orbital independent. Both terms are 
given explicitly in Ref. \cite{Fei99}.

Both the cuprate model (\ref{socu}) and the $e_g$ term in the manganite  
model describe strongly frustrated SE interactions, which 
take a universal form in the limit of $J_H\to 0$,
\begin{equation}
\label{h0e}
{\cal H}_e^{(0)}=\frac{1}{4}J\sum_{\gamma}
                             \sum_{\langle ij\rangle\parallel\gamma}
    \Big[ \Big(\frac{1}{S^2}{\vec S}_i\cdot {\vec S}_j+1\Big)
          \Big(\frac{1}{2}-\tau^{\gamma}_i\Big)
	  \Big(\frac{1}{2}-\tau^{\gamma}_j\Big) - 1 \Big].
\end{equation}
Several classical phases have the same energy of $-3J/4$ per site 
\cite{Fei97}: the $G$-AF phases with arbitrary
occupation of orbitals, and $A$-AF phases with $\langle
(1/2-\tau^{\gamma}_i)(1/2-\tau^{\gamma}_j)\rangle=0$, as obtained for
staggered planar orbitals, e.g. for $x^2-y^2/y^2-z^2$ orbitals staggered 
in $(a,b)$ planes. We emphasize that the model (\ref{h0e}) is 
{\it qualitatively different\/} from the idealized SU(4) symmetric case 
(\ref{su4}) due to the directionality of $e_g$ orbitals. In fact, the 
$e_g$ orbitals order easier, may couple to the lattice and thus appear 
to be more classical than the isotropic case described by Eq. 
(\ref{soiso}). Their ordering supports magnetic phases with coexisting
FM [in $(a,b)$ planes] and AF (along $c$ axis) interactions. 

\begin{figure}[t!]
\begin{tindent}
\epsfxsize=12cm
\epsffile{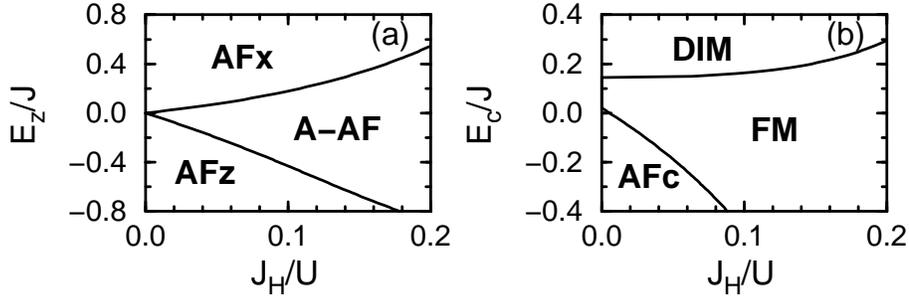}
\end{tindent}
\caption{
Mean-field phase diagrams for: 
(a) one-hole $d^9$ model in $(J_H,E_z)$ plane, and
(b) one-electron $d^1$ model in $(J_H,E_c)$ plane. In the cuprates (a)
the magnetic interactions are predominantly AF and the order changes 
from a 2D AF$x$ through $A$-AF to a 3D AF$z$ phase, while in the 
titanates (b) the DIM, FM, and AFc phases are stable. }
\label{fig:d1}
\end{figure}

The classical phase diagram of the cuprate model (\ref{socu}) is shown
in Fig. \ref{fig:d1}(a). Quantum corrections to this phase diagram were
discussed in Ref. \cite{Fei97}. They suggest that a spin liquid, 
supported by particular OVB correlations, might 
be realized near the degeneracy of classical phases. Finite $J_H$ 
stabilizes the $A$-AF phase, with staggered two-sublattice orbital order, 
$|i\mu\sigma\rangle=\cos\theta_i|iz\sigma\rangle\pm\sin\theta_i|ix\sigma
\rangle$, where $\pm$ refers to $i\in A(B)$ sublattice. The AF 
interactions decrease with increasing $J_H/U$ and dominate at realistic 
$J_H/U\simeq 0.12$ \cite{Miz96}. However, the FM interactions within the 
$(a,b)$ planes predicted by the model \cite{Ole01} are considerably 
stronger than measured \cite{Ten00}, showing that the quantitative 
understanding requires also Goodenough processes which would provide 
additional AF interactions. In contrast, the FM interactions are 
stronger than AF ones in the $A$-AF phase realized in LaMnO$_3$
\cite{Mou99}, and are much better reproduced by the SE terms derived
in Ref. \cite{Fei99} with $\eta=0.117$ \cite{Miz96}.      

The transition metal oxides with partly filled $t_{2g}$ orbitals exhibit 
different and even more interesting phenomena. In this case the JT 
coupling is much weaker, and (unlike for the $e_g$ orbitals) the orbital 
quantum number is conserved in the hopping processes. This leads to 
qualitatively different physics realized in $t_{2g}$ systems,
somewhat more similar to the isotropic case, Eq. (2). 
Each $t_{2g}$ orbital is orthogonal to one of the cubic axes, so we 
label them as $a$, $b$ , and $c$ (for instance, $xy$ orbitals are 
labelled as $c$). The models for titanates and vanadates follow from the 
$d_i^nd_j^n\rightleftharpoons d_i^{n-1}d_j^{n+1}$ processes 
\cite{Kha00,Kha01}, and may be written in a general form:
\begin{equation}
\label{soti}
{\cal H}(d^n)=J\sum_{\gamma}\sum_{\langle ij\rangle\parallel\gamma}
    \Big[({\vec S}_i\cdot {\vec S}_j+S^2)
    {\hat J}_{ij}^{(\gamma)}(d^n)+{\hat K}_{ij}^{(\gamma)}(d^n)\Big],
\end{equation}
with the exchange constants ${\hat J}_{ij}^{(\gamma)}(d^n)$ between 
$S=1/2$ spins for titanates ($n=1$) and $S=1$ spins for vanadates
($n=2$), and purely orbital interactions ${\hat K}_{ij}^{(\gamma)}(d^n)$. 
In titanates these interactions depend on the Hund's rule splittings of 
$d^2$ ions \cite{Gri71} [Fig. \ref{fig:ex}(b)] via the coefficients: 
$r_1=1-3\eta$, $r_2=1-\eta$, $r_3=1+2\eta$, and were given in Refs. 
\cite{Ole01,Khati}. They are faithfully reproduced with a model 
Hamiltonian containing $U$ and $J_H$ for $t_{2g}$ orbitals. 
{\it A~priori,\/} the magnetic interactions are anisotropic, and may be 
either AF or FM, depending on the orbital correlations. 

In the limit of $J_H/U=0$ the Hamiltonian (\ref{soti}) takes the form,
\begin{equation}
\label{h0t}
{\cal H}^{(0)}=\frac{1}{2}
   J\sum_{\gamma}\sum_{\langle ij\rangle\parallel\gamma}
    \Big[ \Big(\frac{1}{S^2}{\vec S}_i\cdot {\vec S}_j+1\Big)
      \Big({\bf \tau}_i\!\cdot\!{\bf \tau}_j+\frac{1}{4}n_in_j\Big)
      -\frac{4}{3}S\Big],
\end{equation}
and shows again a strong frustration of SE interactions \cite{Kha00}.
Although it resembles formally the SU(4)-symmetric spin-orbital models
\cite{Li98} even more than Eq. (\ref{h0e}), the pseudospin operators
${\vec\tau}_i=\{\tau_i^x,\tau_i^y,\tau_i^z\}$ have here a different 
meaning and refer to a pair of orbital flavors for each cubic direction 
$\gamma$, given by two active $t_{2g}$ orbitals which contribute to
the SE \cite{Kha00,Kha01}. Thus, the model is again different from the
idealized SU(4) symmetry Eq. (\ref{su4}).

At finite $\eta$ the magnetic interactions are {\it a priori\/} 
anisotropic, and may be either AF or FM, depending on the orbital 
correlations. In order to get some qualitative insight into the 
competition between these terms, one may include an anisotropy term
$\sim E_c[n_{ic}-(n_{ia}+n_{ib})/2]$, and evaluate the energy of a 
DIM phase with fluctuating $a$ and $b$ orbitals along $c$ axis 
($n_{ia}+n_{ib}=1$), a FM phase with equally and randomly occupied 
orbitals ($n_{i\gamma}=1/3$), and a two-dimensional (2D) AF phase with 
only $c$ orbitals occupied ($n_{ic}=1$). Unlike for $d^9$ case, the FM 
phase is stable in a broad regime of parameters [Fig. \ref{fig:d1}(b)], 
and the DIM phase is stabilized by $E_c\simeq 0.2J$. Of course, this 
analysis is oversimplified and large corrections due to quantum effects 
are expected. Indeed, a FM isotropic phase is realized in YTiO$_3$, but 
a closer inspection shows that the orbitals do order, but 
this ordering does not break the cubic symmetry \cite{Kha02}.
A completely different state is realized in LaTiO$_3$, however, with 
isotropic AF interactions \cite{Kei00}; such interactions are explained
by quantum resonance realized simultaneously in spin and orbital sector 
\cite{Kha00}. This shows that a particular type of magnetic ordering 
may be triggered by quantum fluctuations in the orbital liquid.  

\begin{figure}[t!]
\begin{tindent}
\epsfxsize=12cm
\epsffile{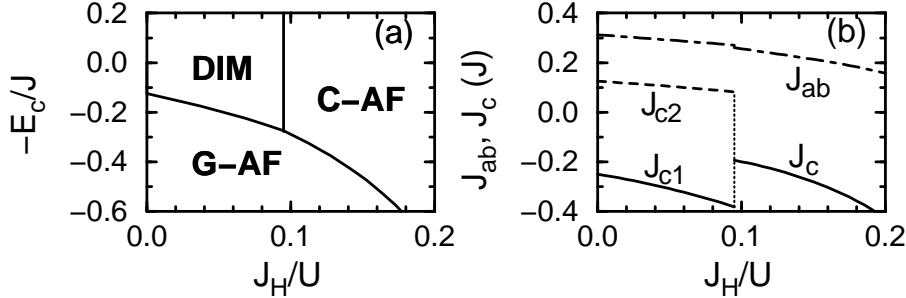}
\end{tindent}
\caption{
(a) Mean-field phase diagram in $(J_H,E_c)$ plane as obtained for the 
vanadate model of Ref. \protect\cite{Kha01}. 
(b) Exchange constants for increasing $J_H$ at $E_c=0$ for the DIM
($J_{c1}$, $J_{c2}$, $J_{ab}$) and for $C$-AF ($J_c$ and $J_{ab}$) 
phase \protect\cite{Hor02}.
}
\label{fig:d2}
\end{figure}

The magnetic ordering realized in vanadates is different: $C$-type of AF 
order is observed both in LaVO$_3$ \cite{Miy00} and in YVO$_3$ at 
intermediate temperatures $77<T<116$ K, and $G$-type AF order is stable 
in YVO$_3$ for $T<77$ \cite{Ren00}. As in V$_2$O$_3$ \cite{Mat02}, the 
SE interactions between $S=1$ spins follow from the 
$d_i^2d_j^2\rightleftharpoons d_i^1d_j^3$ processes, leading to the 
effective spin-orbital model given by Eq. (\ref{soti}) with $n=2$. When 
the electrons condense in $c$ orbitals ($n_{ic}=1$) due to the orbital 
splitting caused by the JT effect, the second electron occupies either 
$a$ or $b$ orbital at every site ($n_{ia}+n_{ib}=1$), allowing for a 
resonance on the bonds $\langle ij\rangle$ along $c$ axis. In this case 
the pseudospin operators are:
$\tau_i^+=a_i^{\dagger}b_i^{}$, $\tau_i^-=b_i^{\dagger}a_i^{}$,
$\tau_i^z=\frac{1}{2}(n_{ia}^{}-n_{ib}^{})$, and
$n_i^{(c)}=n_{ia}^{}+n_{ib}^{}$, where $\{a_i^{\dagger},b_i^{\dagger}\}$
are Schwinger bosons for $a$ and $b$ orbitals at site $i$.

The vanadate model has again an interesting classical phase diagram
which unifies certain features we have already seen for degenerate 
isotropic orbitals, and in the $t_{2g}^1$ model [Fig. \ref{fig:d2}(a)].  
At $\eta=0$ and $E_c=0$ one finds again the frustrated SE (\ref{h0t})
between $S=1$ spins. While the orbital liquid cannot stabilize in this 
case, orbital singlets may form along the $c$ direction when $c$ 
orbitals have condensed ($n_{ic}=1$) and the $a$ and $b$ orbitals 
fluctuate. This favors the OVB state, with strong FM interactions 
alternating with weak AF ones along the one-dimensional (1D) chains 
\cite{She02}. At large $J_H$ this state is unstable, however, and the
orbital fluctuations support FM interactions along $c$ axis and 
stabilize the $C$-AF phase \cite{Kha01}. By considering the energy in 
the MFA one finds a phase transition from the DIM to $C$-AF phase at 
$\eta_c\simeq 0.09$ \cite{Hor02}, as long as $n_{ic}=1$. At large 
uniform orbital splitting $E_c>0$, the charge gets redistributed to
$n_{ia}=n_{ib}=1$, and an anisotropic $G$-AF state with the strong AF
bonds along $c$ axis, and weaker ones within $(a,b)$ planes, follows.
To our knowledge, such a state has not been observed so far. The $G$-AF 
phase found in YVO$_3$ at $T<T_{N1}$ is characterized by large JT
distortions, and thus can be explained by a staggered field which 
favors $C$-type orbital ordering \cite{Kha01}, in agreement with recent 
experiments \cite{Ren00}.

The SE interactions in transition metal oxides depend on the multiplet 
splittings $\propto J_H$. For example, if $n_{ic}=1$ and 
$n_{ia}+n_{ib}=1$ in cubic vanadates, the exchange constants within the 
$(a,b)$ planes ($J_{ab}$) and along the $c$ axis ($J_c$) in the $C$-AF 
phase are given by \cite{Kha01}:
\begin{eqnarray}
\label{jab}
J_{ab}&\!=&\![1-\eta (R\!+\!r)
 +(1+2\eta R-\eta r)\langle n_{ia}n_{ja}\rangle^{(b)}]/4,      \\
\label{jc}
J_c&\!=&\![(1\!+\!2\eta R)
 \langle{\vec\tau}_i\!\cdot\!{\vec\tau}_j+1/4\rangle^{(c)}
-\eta r\langle\tau_i^z\tau_j^z+1/4\rangle^{(c)}-\eta R]/2,
\end{eqnarray}
where $R=1/(1-3\eta)$ and $r=1/(1+2\eta)$. 
Similar expressions can be derived in the DIM phase at $\eta<\eta_c$. 
Assuming orbital singlets in the DIM phase, and strong 1D orbital 
fluctuations, described by a pseudospin 1D Heisenberg model in the 
$C$-AF phase, one finds that the FM/AF exchange constants coexist and 
increase/decrease with increasing $J_H$ [Fig. \ref{fig:d2}(b)]. It is 
interesting to observe that the values of $J_{ab}$ and $|J_{c}|$ are 
similar for a realistic value of $\eta\simeq 0.116$ \cite{Miz96}, as 
the orbital fluctuations enhance the FM interactions $\propto J_{c}$.

\begin{figure}[t!]
\begin{tindent}
\epsfxsize=12cm
\epsffile{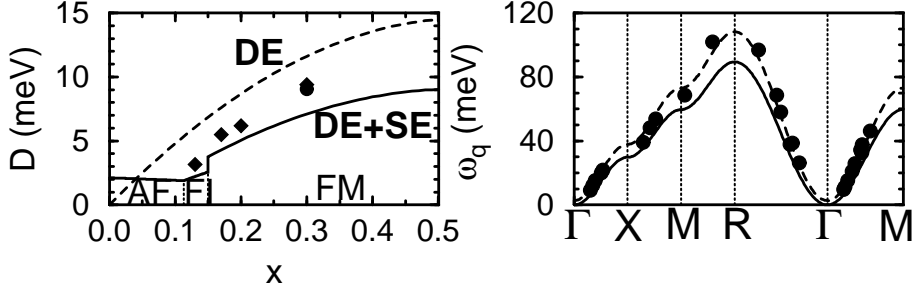}
\end{tindent}
\caption{
(a) Spin-wave stiffness $D$ as a function of hole doping $x$ obtained
from the DE (dashed line), and by including also the SE terms (solid 
line) for: $A$-AF (AF), FI, and FM metallic phase \protect\cite{Ole02}. 
Experimental points for:
La$_{1-x}$Sr$_x$MnO$_3$ \protect\cite{End97} (diamonds) and
La$_{0.7}$Pb$_{0.3}$MnO$_3$ \protect\cite{Per96} (filled circle).
(b) Magnon dispersion $\omega_{\vec q}$ for $x=0.30$ (solid line) 
\protect\cite{Ole02}, and the data points for 
La$_{0.7}$Pb$_{0.3}$MnO$_3$ (circles and dashed line) 
\protect\cite{Per96}.
}
\label{fig:mag}
\end{figure}

Let us come back to the question why the orbital liquid state cannot
stabilize in LaMnO$_3$. In this case the orbitals do order, and the 
orbital interactions are so strong that their ordering 
would occur well above $T_N$ even in the absence of the JT interaction
\cite{Fei99}. However, the splitting between the high-spin $^6A_1$ state 
and low-spin states is $5J_H$ (Fig. \ref{fig:ex}a), which explains the 
proximity to the FM ordering. The manganites at $x<0.15$ are insulating 
\cite{Mou99}, and are orbital ordered, with either $A$-AF or FM 
insulating (FI) phase due to polaronic effects [Fig. \ref{fig:mag}(a)].
Therefore, a single hole does not propagate freely but scatters on 
orbital \cite{vdB00} and spin \cite{Bal01} excitations. We discussed 
elsewhere that doping $x>0.15$ stabilizes the FM metallic state due to 
the double exchange (DE) for strongly correlated $e_g$ orbitals 
\cite{Ole02}. This FM metallic state is nothing else than the 
realization of the orbital liquid in an $e_g$ system. By considering the 
DE and SE together one arrives at a quantitative explanation of: 
 $(i)$ the spin-wave stiffness $D$ increasing with $x$ \cite{End97},
       which just reflects the gradual release of the kinetic energy by 
       hole doping [Fig. \ref{fig:mag}(b)];  
$(ii)$ the {\it isotropic\/} spin waves observed around $x=0.3$ in a 
       system being so susceptible towards the orbital ordering.
The spin-wave stiffness $D_{\rm eff}=7.45$ meV obtained at $x=0.3$ 
without any fitting parameters agrees well with $D_{\rm exp}=8.79$ meV
measured in La$_{0.7}$Pb$_{0.3}$MnO$_3$ \cite{Per96} and explains the 
observed dispersion $\omega_{\vec q}$ [Fig. \ref{fig:mag}(b)]. A
transition from the FM to $A$-AF phase observed in bilayer manganites
La$_{2-2x}$Sr$_{1+2x}$Mn$_2$O$_7$ \cite{Per01} can also be explained
within the same approach.

In summary, the transition metal oxides with orbital degrees of freedom
show a very fascinating behavior, with various types of {\it magnetic 
and orbital order\/}. While $e_g$ orbitals usually order and explain
$A$-AF phases, further stabilized by the JT effect, the $t_{2g}$
orbitals have a generic tendency towards disorder, which leads to the
{\it isotropic\/} orbital liquid in the $G$-AF phase in LaTiO$_3$, and 
to a 1D {\it anisotropic\/} orbital liquid in the $C$-AF phase in 
LaVO$_3$ and YVO$_3$. So strong orbital fluctuations in $e_g$ systems 
and the orbital liquid state are triggered only by large doping in the 
manganites. Very interesting quantum effects might also soon be 
discovered in the orbital liquid states in doped titanates and 
vanadates.

{\it Acknowledgments.\/}
It is a pleasure to thank Lou-Fe' Feiner, Peter Horsch, Giniyat
Khaliullin, and Jan Zaanen for a very friendly collaboration on 
this subject and for numerous stimulating discussions. I thank 
also J. Ba\l{}a, A. Fujimori, B. Keimer, G.A. Sawatzky, and 
Y. Tokura for valuable discussions. 
This work was supported by the Committee of Scientific Research (KBN)
of Poland, Project No. 2~P03B~055~20.


\end{document}